\documentclass[a4paper,11pt]{article}
\usepackage{pos,enumitem,float,pifont}
\usepackage{xcolor, soul}

\usepackage{tikz}
\usetikzlibrary{arrows.meta}
\usetikzlibrary{decorations.markings}
\tikzstyle midarrow=[postaction={decorate,decoration={markings,
      mark=at position 0.53 with {\arrow{#1}},}}]

\graphicspath{{Figures/}}

\title{$B$ meson Decay Constants Using Relativistic Heavy Quarks}

\author*{Matthew Black}
\author{Oliver Witzel}

\affiliation{\textbf{(RBC and UKQCD collaborations)}\\
    Center for Particle Physics Siegen, Theoretische Physik 1,
  Naturwissenschaftlich-Technische Fakult\"at,
  Universit\"at Siegen, 57068 Siegen, Germany}

\emailAdd{matthew.black@uni-siegen.de}

\abstract{We present an update on ongoing work to extract pseudoscalar and vector decay constants for $B^{(*)}$, $B^{(*)}_s$ and $B^{(*)}_c$ mesons and determine phenomenologically-interesting ratios such as $f_{B_s}/f_B$ or $f_{B^*}/f_B$.
Our calculation is based on ${\rm N_f}=2+1$ dynamical flavour gauge field ensembles generated by the RBC/UKQCD collaborations using domain-wall fermions and the Iwasaki gauge action.
Using domain-wall light, strange, and charm quarks and relativistic $b$ quarks, we obtain results at multiple lattice spacings and valence quark masses.}

\FullConference{%
  The 39th International Symposium on Lattice Field Theory,\\
  8th-13th August, 2022,\\
  Rheinische Friedrich-Wilhelms-Universität Bonn, Bonn, Germany
}

\def\gev{\,\text{Ge\hspace{-0.1em}V}}
\def\mev{\,\text{Me\hspace{-0.1em}V}}
\def\fm{\,\text{fm}}

\begin{document}
\maketitle

\section{Introduction}
Leptonic decays of $B$ mesons are some of the simplest flavour-changing processes in which a $b$ quark can participate. 
In the Standard Model (SM), the decay rate of a $B^+$ meson decaying into a charged-lepton--neutrino pair is given by
\begin{equation}
    \label{eq:GammaB}
    \Gamma(B^+\to\ell^+\nu_\ell) = \frac{G_F^2m_Bm_\ell^2}{8\pi}\left(1-\frac{m_\ell^2}{m_B^2}\right)^2\,|V_{ub}|^2\,f_B^2.
\end{equation}
From the perspective of QCD, this is a very clean process where all of the nonperturbative, hadronic physics is contained within the decay constant $f_B$.
The simplicity of Eq.\eqref{eq:GammaB} would motivate using these decay rates to determine the CKM matrix element $|V_{ub}|$, however this requires high precision knowledge of both the measured decay rate and the decay constant $f_B$. 
The limiting factor in this respect is still experiment where thus far $B^+\to\tau^+\nu_\tau$ and $B^+\to\mu^+\nu_\mu$ have been observed with precision $\sim 30\%$ and $40\%$ respectively \cite{Aubert:2009wt,Belle:2010xzn,Lees:2012ju,Adachi:2012mm,Belle:2015odw,Belle:2019iji}; this precision is anticipated to be improved upon in the near-future using the new data being collected by Belle II.
The pseudoscalar decay constants of $B$ mesons with a light/strange valence quark are well-studied using nonperturbative methods. 
From QCD sum rules, the best determinations have a precision $\sim10\%$ \cite{Lucha:2010ea,Gelhausen:2013wia,Pullin:2021ebn}. 
With lattice QCD, these are currently known to a precision of less than $1\%$ \cite{Aoki:2021kgd,Bazavov:2017lyh,ETM:2016nbo,Dowdall:2013tga,Hughes:2017spc}; from the pseudoscalar decay constants also comes the important SU(3)-breaking ratio $f_{B_s}/f_{B}$.
Furthermore, these decay constants are important in other processes such as in extracting bag parameters of neutral $B$-meson mixing from hadronic matrix elements or in predicting the rare leptonic decays $B_q\to\ell^+\ell^-$.

As well studied as the pseudoscalar decay constants are, one can also consider the decay constants of more `exotic' $B$ states. 
As the next generation of experiments begin, such as HL-LHC and Belle II, and further theoretical techniques allow the community to predict more `exotic' decay channels, there is more motivation than ever to develop high-precision lattice calculations of new decay constants, with lower-hanging `exotic' fruits being for example the heavier $B_c$ meson and the vector states $B_q^*,\,q= l,s,c$. 
To date, only one calculation of the pseudoscalar $B_c$ decay constant has been performed by the HPQCD collaboration \cite{Colquhoun:2015oha}.
Similarly, there exists only two lattice calculations of the decay constant for the vector meson state $B_q^*$ from the HPQCD \cite{Colquhoun:2015oha} and ETM \cite{Becirevic:2014kaa} collaborations. 
Of additional interest in the case of the vector decay constant is that these two calculations (with $q=l,s$) are in tension with one another and in fact their ratios $f_{B^*_{(s)}}/f_{B_{(s)}}$ lie to opposite sides of unity.
The single calculation of $f_{B_c}$ and the tension in $f_{B^*_{(s)}}$ motivates further lattice calculations of these quantities to first reach consensus on their values and then push to higher precision. 
We will contribute to this by analysing the pseudoscalar and vector decay constants for $B,\,B_s,$ and $B_c$ mesons.

\section{Lattice calculation}
We use seven RBC/UKQCD $2{+}1$-flavour domain-wall fermion (DWF) and Iwasaki gauge field ensembles with four lattice spacings $a\sim 0.11$, $0.08$, $0.07$, $0.06\fm$ (determined from RBC/UKQCD analyses~\cite{Blum:2014tka,Boyle:2017jwu,Boyle:2018knm}) and pion masses $\in[267,433)\mev$. 
Light and strange quarks are simulated with the Shamir DWF action with $M_5=1.8$. 
These ensembles are listed in Table~\ref{tab:ensembles}. 
\begin{table}[th]
\[
  \begin{array}{cccccccccccc}
  & L & T & L_s &  a^{-1}\!/\!\gev & am_l^\text{sea} & am_s^\text{sea} 
  & M_\pi/\!\mev & \text{\# cfgs} & \text{\# sources}\\\hline
\text{C1} & 24 & 64 & 16 & 1.785 & 0.005 & 0.040 & 340 & 1636 & 1\\
\text{C2} & 24 & 64 & 16 & 1.785 & 0.010 & 0.040 & 433 & 1419 & 1\\[1.2ex]
\text{M1} & 32 & 64 & 16 & 2.383 & 0.004 & 0.030 & 302 & 628  & 2\\
\text{M2} & 32 & 64 & 16 & 2.383 & 0.006 & 0.030 & 362 & 889  & 2\\
\text{M3} & 32 & 64 & 16 & 2.383 & 0.008 & 0.030 & 411 & 544  & 2\\[1.2ex]
\text{F1S}& 48 & 96 & 12 & 2.785 & 0.002144 & 0.02144 & 267 &98 & 24 \\[1.2ex]
\text{X1} & 32 & 64 & 12 & 3.148 & 0.0047 & 0.0186 & 371 & 40 & 16
  \end{array}
  \]
  \caption{Ensembles used for the simulations reported here~\cite{Allton:2008pn,Aoki:2010dy,Blum:2014tka,Boyle:2017jwu}.
    $am_l^\text{sea}$ and $am_s^\text{sea}$ are the light and strange sea quark masses and $M_\pi$ is the unitary pion mass. 
    Valence strange quarks are near their physical mass.}
  \label{tab:ensembles}
\end{table}
  
As bottom quarks are too heavy to be simulated directly at their physical value on these lattices, we choose the effective action approach in order to simulate physical $b$ quarks. 
Specifically, we choose the relativistic heavy quark (RHQ) action, which is the Columbia variant~\cite{Christ:2006us,Lin:2006ur} of the Fermilab heavy-quark action~\cite{ElKhadra:1996mp}, with three nonperturbatively-tuned parameters $m_0a,\,c_P,$ and $\zeta$~\cite{Aoki:2012xaa}. 
The tuning of these parameters on each ensemble was initially carried out in Ref.~\cite{Aoki:2012xaa}; we report here first results for tuning the RHQ parameters on the X1 ensemble.
Charm quarks are simulated with the M\"obius DWF action and $M_5=1.6$~\cite{Boyle:2016imm,Boyle:2017jwu,Boyle:2017kli,Boyle:2018knm}.
We use three masses below $m_c^\text{phys}$ on C and two masses bracketing $m_c^\text{phys}$ on M, F, and X. 
Light and strange quarks have point sources, while the $b$ and $c$ quarks use Gaussian-smeared sources and point or smeared sinks; to improve statistics on X, instead of point sources we use Z2-noise wall sources which are also Gaussian-smeared for the $b$ quarks.

Here we consider the vacuum-to-meson matrix elements of $B_q$ mesons at rest (where $q=l,s,c$ denotes the valence quark) $\langle 0 | J_\mu | B_q(0) \rangle$ for the axial-vector $A_\mu$ and vector $V_\mu$ currents to yield the decay constants of the pseudoscalar and vector meson states respectively, where for the pseudoscalar we require only the temporal component $A_0$ and for the vector the spatial components $V_i,\,i=1,2,3$.
For these matrix elements, we can relate a lattice current $J_\mu$ to the renormalised current $\mathcal{J}_\mu$ via the multiplicative renormalisation factor 
\begin{equation}
    Z_{J_\mu}^{bq} = \rho_{J_\mu}^{bq}\sqrt{Z^{qq}_V\, Z^{bb}_V},
\end{equation}
which follows the partially-nonperturbative approach~\cite{Hashimoto:1999yp,ElKhadra:2001rv} where $Z_V^{bb}$ and $Z_V^{qq}$ are nonperturbatively calculated and $\rho_{J_\mu}^{bq}$ is the remaining correction calculated perturbatively at one-loop and is expected to only deviate from unity by ${\cal O}(1\%)$ \cite{Christ:2014uea,CLehnerPT}.
For DWFs, $Z_V^{qq}$ can be computed via the relation
\begin{equation}
    Z_V^{qq} = Z_A^{qq} + {\cal O}(am_{\rm res}),
\end{equation}
and $Z_V^{bb}$ is calculated from the matrix element
\begin{equation}
    Z_V^{bb}\langle B_q | V_0 | B_q \rangle = 2M_{B_q}.
\end{equation}

Using the RHQ action, in order to reduce cut-off effects we require for lattice (axial-)vector currents to include ${\cal O}(a)$-improvement terms before this renormalisation. 
For the spatial components of the vector current $V_i = \bar{b}\gamma_i q$, the relevant subset of the ${\cal O}(a)$-improvement operators for calculating the decay constant are
\begin{align}
    V_i^0 &= \bar{b}\gamma_i q, &
    V_i^2 &= \bar{b}(2\overleftarrow{D}_i)q, &
    V_i^4 &= \bar{b}\gamma_i\sum_j\gamma_j(2\overleftarrow{D}_j)\,q,
\end{align}
and similarly for the axial-vector current $A_0 = \bar{b}\gamma_0\gamma_5q$ replacing $\gamma_i\to\gamma_0\gamma_5$. 
By considering only the temporal component, $A_0^2$ is eliminated using the equations of motion.
Then, we can write the required ${\cal O}(a)$-improved currents as
\begin{align}
    A_0 &= A_0^0 + c_A^4\, A_0^4, \\
    V_i &= V_i^0 + c_V^2 V_i^2 + c_V^4 V_i^4.
\end{align}
The coefficients $c_{A,V}^k$ of the ${\cal O}(a)$-improvement operators are computed at one-loop perturbation theory~\cite{CLehnerPT,Christ:2014uea}.

\begin{figure}
    \centering
    \includegraphics[width=0.5\textwidth]{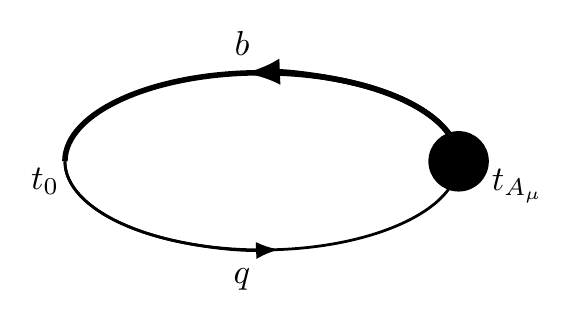}
    \caption{Two-point correlator used to determine a pseudoscalar decay amplitude. $t_0$ indicates the position of the source with pseudoscalar current $\bar{b}\gamma_5q$, and $t_{A_\mu}$ is the time of insertion of the axial-vector current $A_\mu$.}
    \label{fig:2ptcorr}
\end{figure}

The renormalised vacuum-to-meson matrix elements are related to the decay constants as
\begin{align}
    \langle 0 | {\cal A}_0 | B_q(0) \rangle &= f_{B_q}M_{B_q}, \\
    \langle 0 | {\cal V}_i | B_{q,j}^*(0) \rangle &= f_{B_q^*}M_{B_q^*}\delta_{ij}.
\end{align}
It can be simpler first to consider the decay amplitude $\Phi_{B_q} = f_{B_q}\sqrt{M_{B_q}}$ which can be defined as a ratio of two-point correlators in the limit of $t_0\ll t\ll T+t_0$ such that excited states have decayed and the ground state of the system is clearly resolved.
By `folding' correlators such that they only have forwards-propagating states up to extent $T/2$ and $t_0=0$, we need only consider the limit $t\gg0$ to find the ground state.
For pseudoscalar and vector decay amplitudes respectively, these ratios are defined as
\begin{align}
    \label{eq:PhiPS}
    \Phi_{B_q} &= \sqrt{2}\,Z_{A_0}^{bq}\; \lim_{t\gg0} \frac{|\tilde{C}_{A_0P}(t,0)|}{\sqrt{\tilde{\tilde{C}}_{{P}{P}}(t,0)\,e^{-M_{B_q}t}}}, \\
    \Phi_{B_q^*} &= \sqrt{2}\,Z_{V_i}^{bq}\; \lim_{t\gg0} \frac{|\tilde{C}_{V_iV_i}(t,0)|}{\sqrt{\tilde{\tilde{C}}_{V_i^0V_i^0}(t,0)\,e^{-M_{B_q^*}t}}},
\end{align}
where we define $P$ the pseudoscalar current $\bar{b}\gamma_5q$ and a two-point correlation function $C_{XY}(t,t_0)$ with currents $X,Y$ inserted at times $t,t_0$ respectively. $\tilde{C}$ indicates the correlator is smeared at the source, and $\tilde{\tilde{C}}$ indicates both source and sink are smeared. 
A schematic of the two-point correlator $C_{A_0P}(t,t_0)$ is shown in Figure~\ref{fig:2ptcorr}.

\begin{figure}[th]
    \centering
    \includegraphics[width=0.49\textwidth]{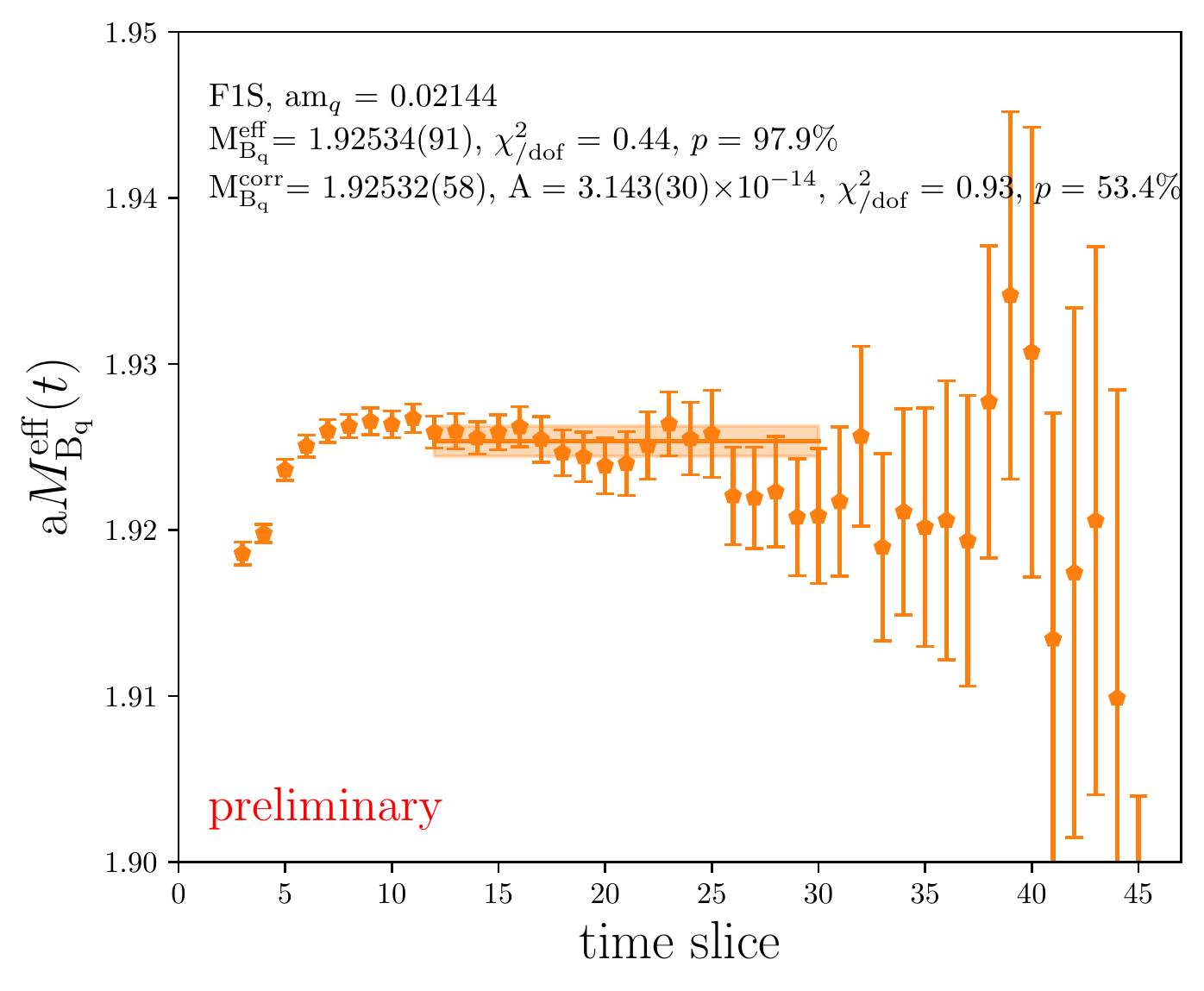}
    \includegraphics[width=0.49\textwidth]{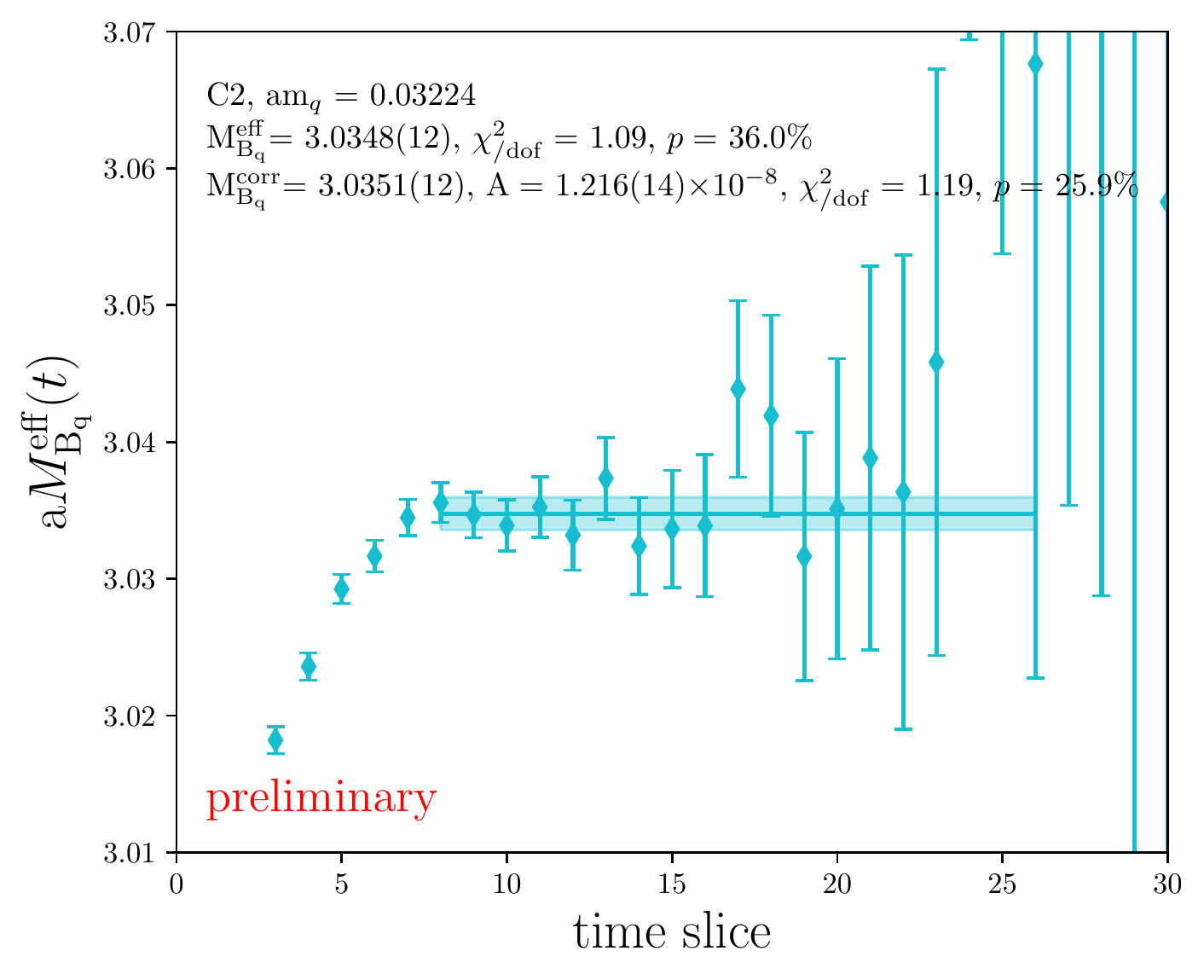}
    \includegraphics[width=0.49\textwidth]{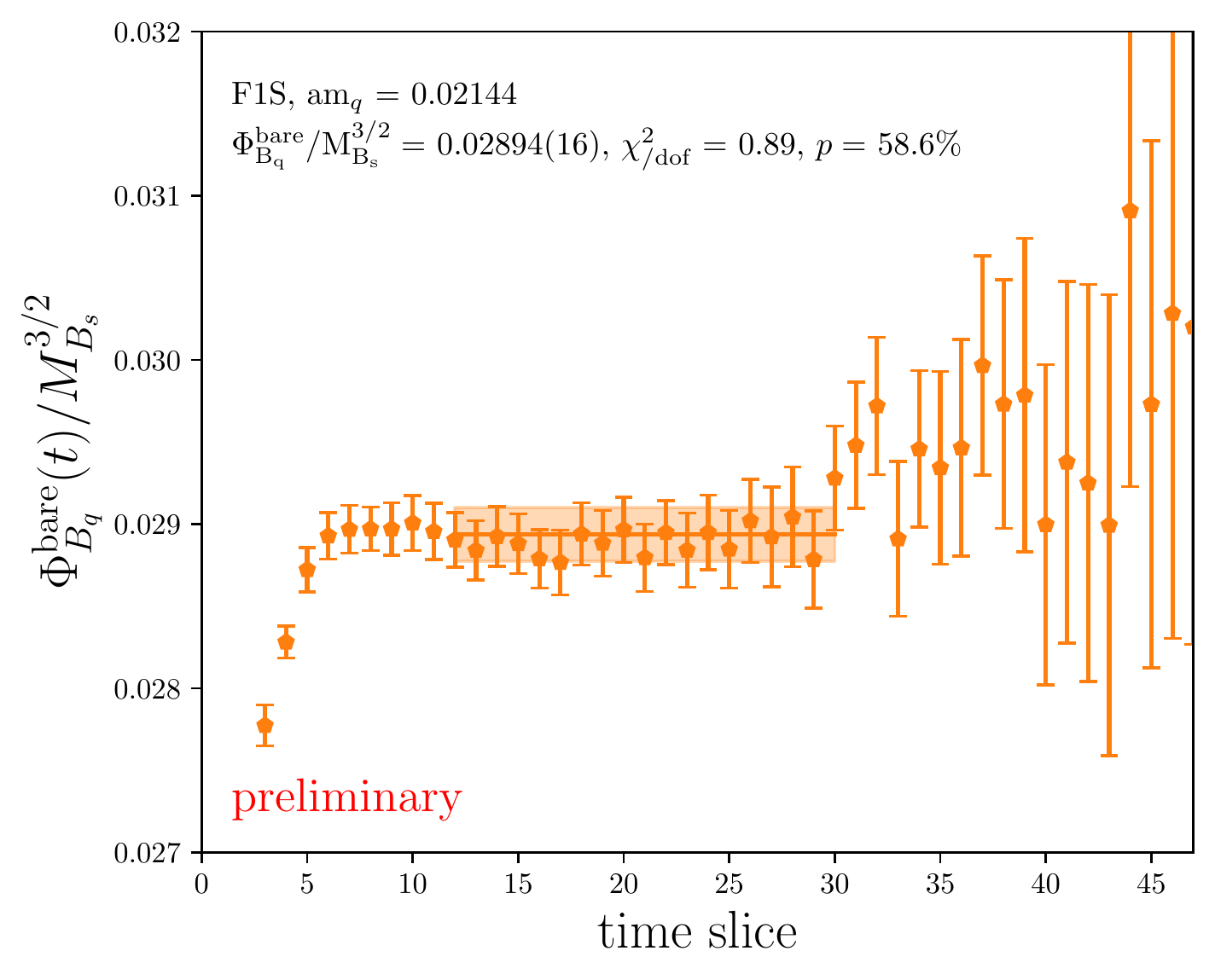}
    \includegraphics[width=0.49\textwidth]{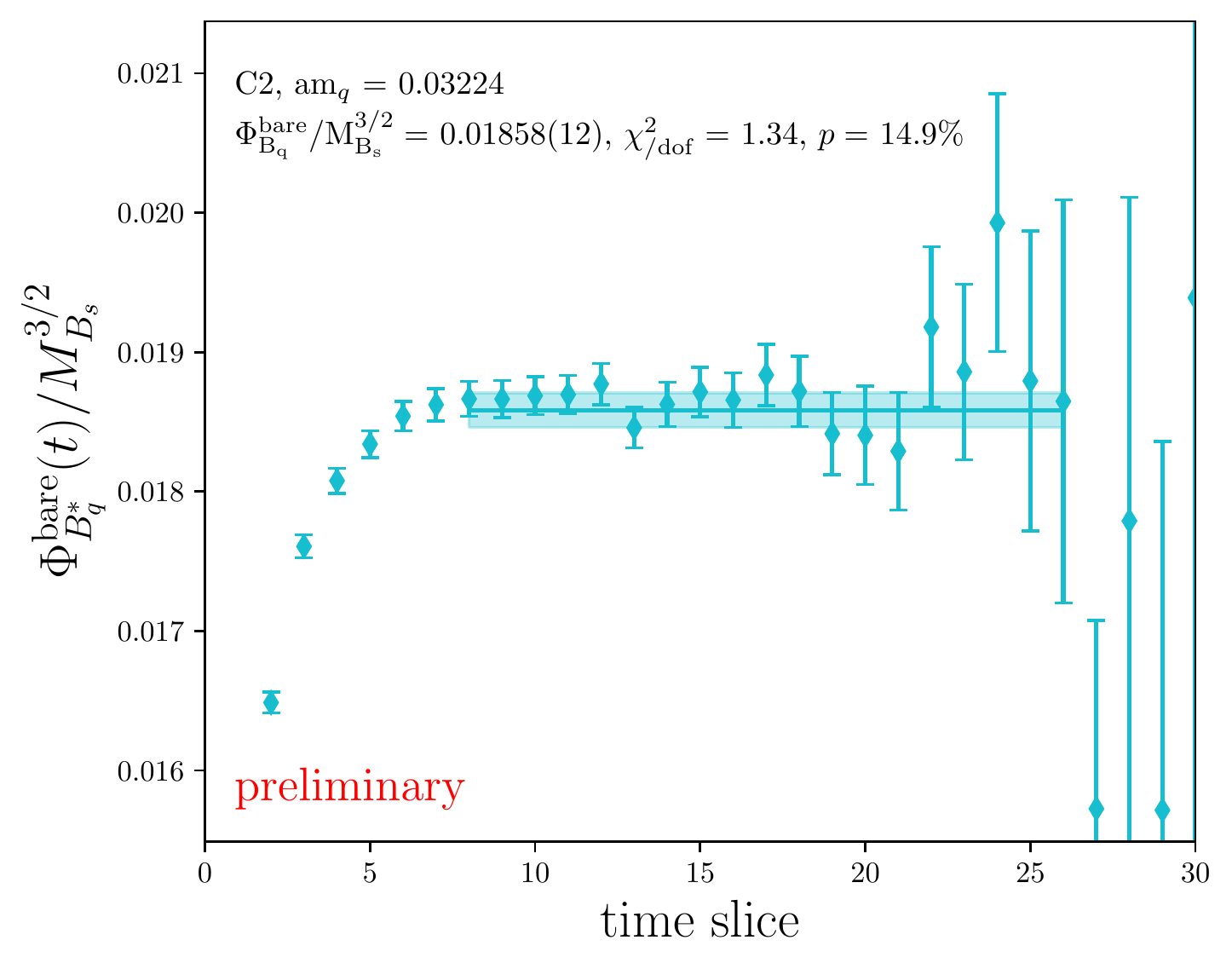}
    \caption{Determination of the effective mass (top) and dimensionless decay amplitude (bottom) for the pseudoscalar on F1S (left) and for the vector on C2 (right). The data are fitted to a constant value in the region where excited state contamination is minimised and we have a clear ground state signal.}
    \label{fig:C1_PhiBM}
\end{figure}

Finally, we present the decay amplitudes as dimensionless ratios over the $B_s$ meson mass $M_{B_s}^{3/2}$, which is used to tune the RHQ parameters to the physical value of the $b$ quark.
To gain our final results in physical units after performing continuum extrapolations, we can simply multiply by the experimental value of $M_{B_s}$ \cite{ParticleDataGroup:2022ynf}.
In Figure~\ref{fig:C1_PhiBM}, we show examples of the effective mass $M_{B_s^{(*)}}$ and dimensionless bare decay amplitude $\Phi_{B_q^{(*)}}/M_{B_s}^{3/2}$.

\section{Chiral and Continuum Extrapolations}
Here we show the current status of our analysis to extrapolate to the physical values for the decay amplitudes of pseudoscalar $B_{(s)}$ systems. 
Analysis of the $B_c$ follows analogously with an extra-/interpolation to physical charm mass using the $\eta_c$ meson mass.
Measurements are still in progress for the vector states and thus extra-/interpolations to physical values cannot yet be performed.

For the determination of $\Phi_{B_s}$ we currently consider a two step procedure. First we correct for slight mistunings of the used strange quark mass by inter-/extrapolating to the physical value. That way we obtain values for $\Phi_{B_s}$ for each of our six different ensembles and use these values to also form SU(3) symmetry breaking ratios $\Phi_{B_s}/\phi_B$. In order to determine $f_{B_s}$ we need to renormalize our $\Phi_{B_s}$ values and then perform in a second step a continuum limit extrapolation in $a^2$.

\begin{figure}[th]
    \centering
    \includegraphics[width=0.49\textwidth]{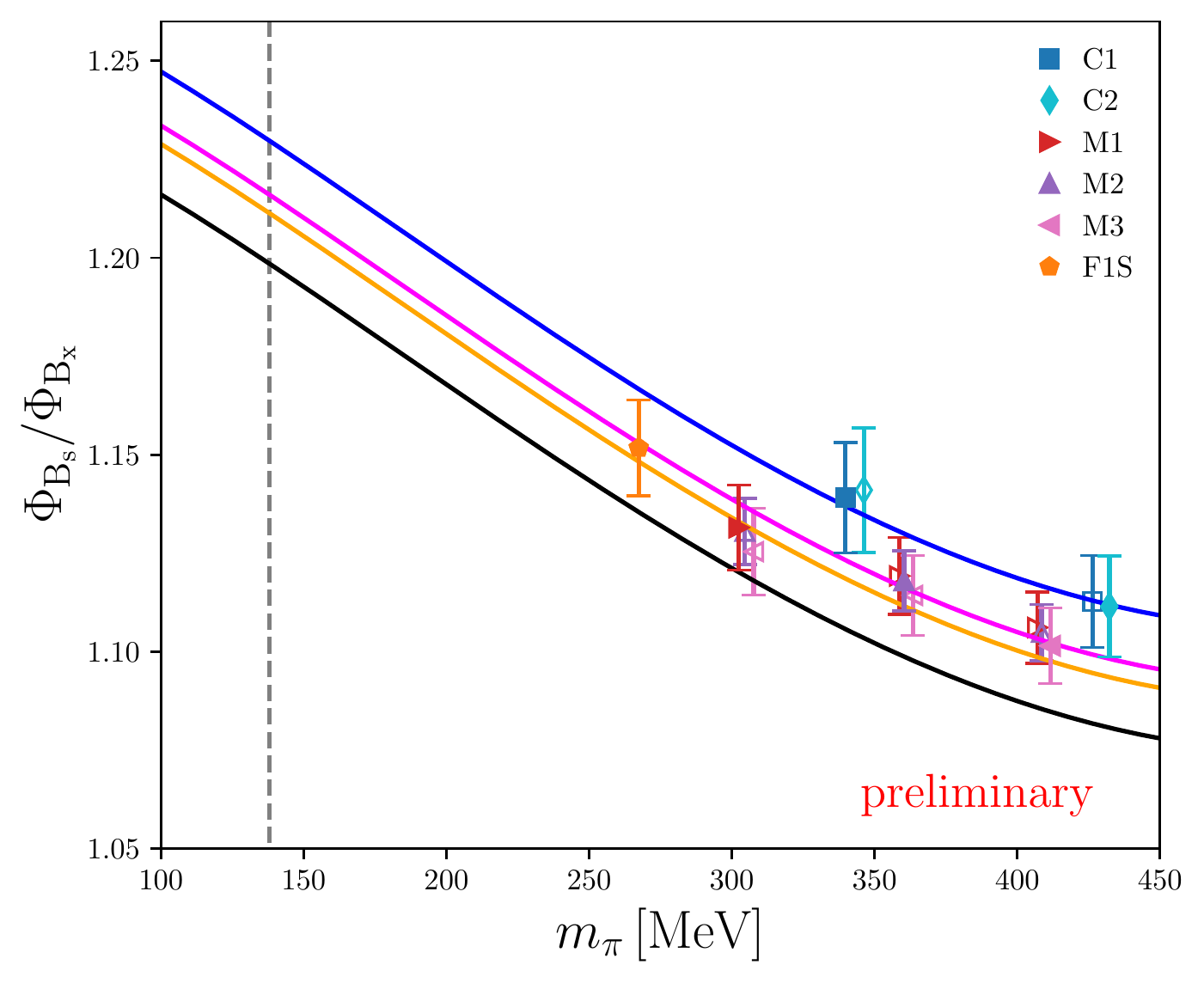}
    \caption{Chiral-continuum extrapolation of $\Phi_{B_s}/\Phi_{B_q}$ using NLO SU(2) HM$\chi$PT. 
        The coloured data are associated with different gauge configurations as indicated in the legend. 
        Only the filled points (unitary) enter the fits. 
        The coloured lines indicate the fit function at each corresponding lattice spacing, the black line indicates the function at zero lattice spacing. 
        The black dashed line indicates the isospin-averaged pion mass, $m_{\bar\pi}\sim138\,$MeV. 
        Only statistical errors are shown.}
    \label{fig:PhiBl}
\end{figure}

Skipping the renormalisation of the decay amplitudes for now, we continue looking at the SU(3) symmetry breaking ratio $\Phi_{B_s}/\Phi_B$ right away. To perform an extrapolation of our data shown in Figure \ref{fig:PhiBl}, we use
NLO SU(2) Heavy Meson Chiral Perturbation Theory (HM$\chi$PT). 
For sufficiently light valence quark masses, this will provide an effective description of the decay amplitude as a function of the pion mass.
This allows us to take the joint continuum and physical pion mass limit from our data. 
Similarly 
we use HM$\chi$PT to also calculate the SU(3)-breaking ratio $f_{B_s}/f_{B}$ 
In the limit of unitary pions, our fit Ans\"atze for these extrapolations are
\begin{align}
    \frac{\Phi_{B_x}}{M_{B_s}^{3/2}} &= \Phi_0 \left\{ 1 - \frac{1+3g_b^2}{(4\pi f_\pi)^2}\frac34 \cdot M_{xx}^2\ln(M_{xx}^2/\Lambda_\chi^2) + c_1\cdot\frac{2M_{xx}^2}{(4\pi f_\pi)^2}+c_2\cdot\frac{a^2}{(4\pi f_\pi)^2}\right\},\\
    \frac{\Phi_{B_s}}{\Phi_{B_x}} &= R_{\Phi}\left\{ 1 + \frac{1+3g_b^2}{(4\pi f_\pi)^2}\frac34 \cdot M_{xx}^2\ln(M_{xx}^2/\Lambda_\chi^2)\right\} + d_1\cdot\frac{2M_{xx}^2}{(4\pi f_\pi)^2}+ d_2\cdot\frac{a^2}{(4\pi f_\pi)^2},
\end{align}
where $\Phi_0,c_1,c_2,R_{\Phi},d_1,d_2$ are fit coefficients, $M_{xx}$ is the mass of a pion with two valence quarks of mass $m_x$, $g_b=0.56(3)$ the $B^*B\pi$ coupling constant \cite{Flynn:2015xna}, and $f_\pi=130.2(1.2)\,$MeV the pion decay constant \cite{ParticleDataGroup:2020ssz}. 
The chiral-continuum extrapolation for $\Phi_{B_s}/\Phi_{B_x}$ is shown in Figure~\ref{fig:PhiBl}.

It is also possible for the above fits to choose different Ans\"atze to perform our extrapolations. 
The variation of results from different Ans\"atze will be included in a full systematic error analysis later on.
Further systematic effects to be considered include, for example, sea quark mass mistuning and RHQ parameter uncertainties.

\section{Tuning RHQ on X1}
In early analysis of the $B_c$ decay constant using C, M, and F ensembles, large cut-off effects are found, particularly on the coarse C ensembles.
Therefore, to improve extrapolation to a physical value, we choose to include the extra-fine ensemble in the RHQ analysis: X1.
To use this additional ensemble, we perform the nonperturbative tuning procedure for the RHQ parameters $m_0,\,c_P,$ and $\zeta$ as previously done for the other ensembles \cite{Aoki:2012xaa}.
Measurements performed on the X1 ensemble make use of \texttt{Grid} \cite{GRID,Boyle:2016lbp} and \texttt{Hadrons} \cite{Hadrons22}.

\begin{figure}[th]
    \centering
    \includegraphics[width=0.49\textwidth]{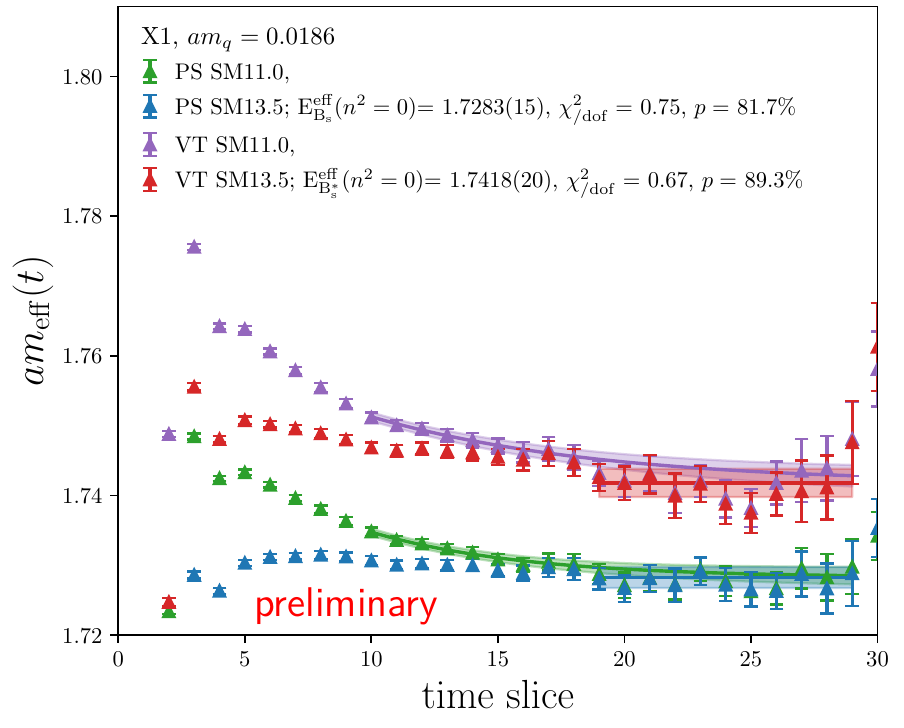}
    \includegraphics[width=0.49\textwidth]{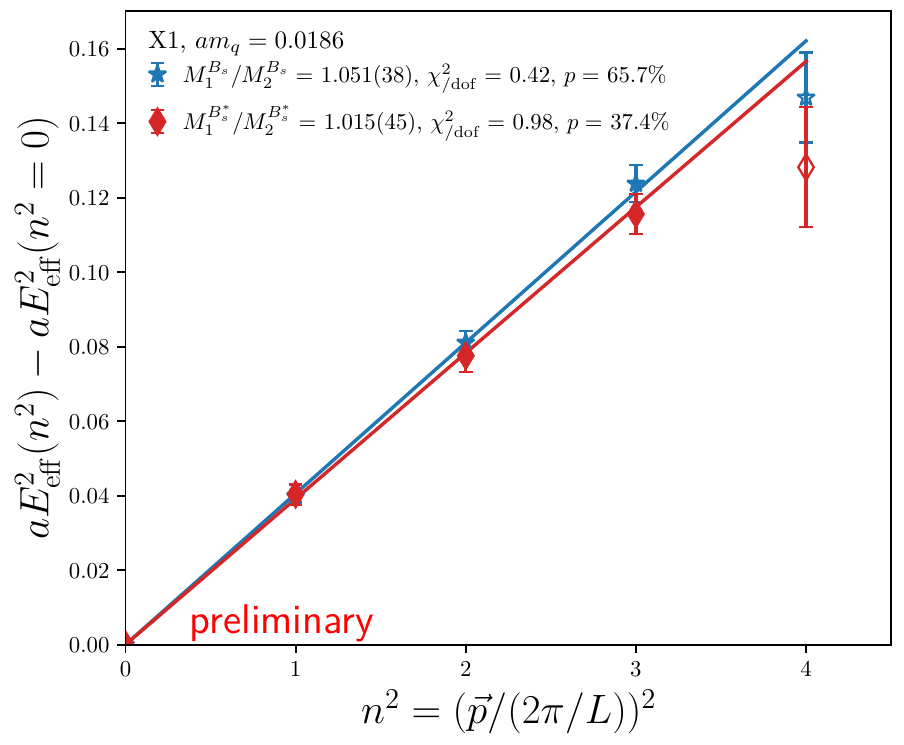}
    \caption{Examples of fits included in the tuning analysis on X1. {\it Left plot:} extraction of the ground state meson mass with $\vec{p}^2=0$ for pseudoscalar $B_s$ and vector $B_s^*$. The valence quark is set to the unitary strange mass, $am_q=0.0186$. Two correlators with different Gaussian-smearing radii are used in a combined fully-correlated fit and a first excited state is also included for the smaller smearing radius. {\it Right plot:} a fit to the slope of the continuum dispersion relation where ground state energies are first extracted for each integer square momentum $n^2$, and we fit to $n^2\leq3$. Tuned RHQ parameters enforce a slope equal to 1.}
    \label{fig:X1Tune}
\end{figure}

This RHQ tuning process is not yet completed for this analysis, however at the current stage of this procedure, we find promising signals of the potential of X1 to aid in the analysis of the $B_c$ decay constants, and potentially the light $B_{(s)}$ states as well. 
Examples of fits entering the tuning analysis are show in Figure~\ref{fig:X1Tune}.

\section{$B_c$ Decay Constant}
\begin{figure}[th]
    \centering
    \includegraphics[width=0.50\textwidth]{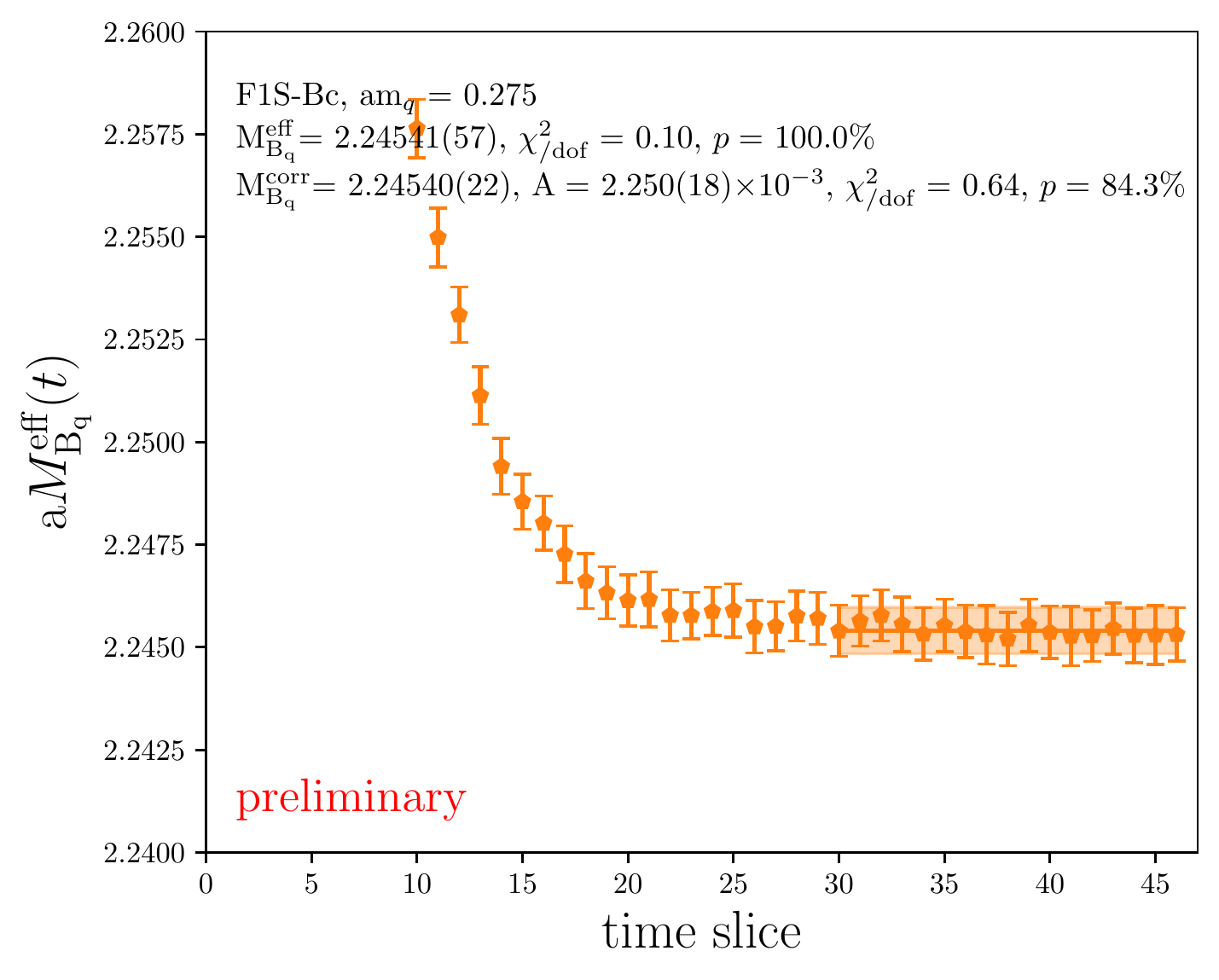}
    \includegraphics[width=0.465\textwidth]{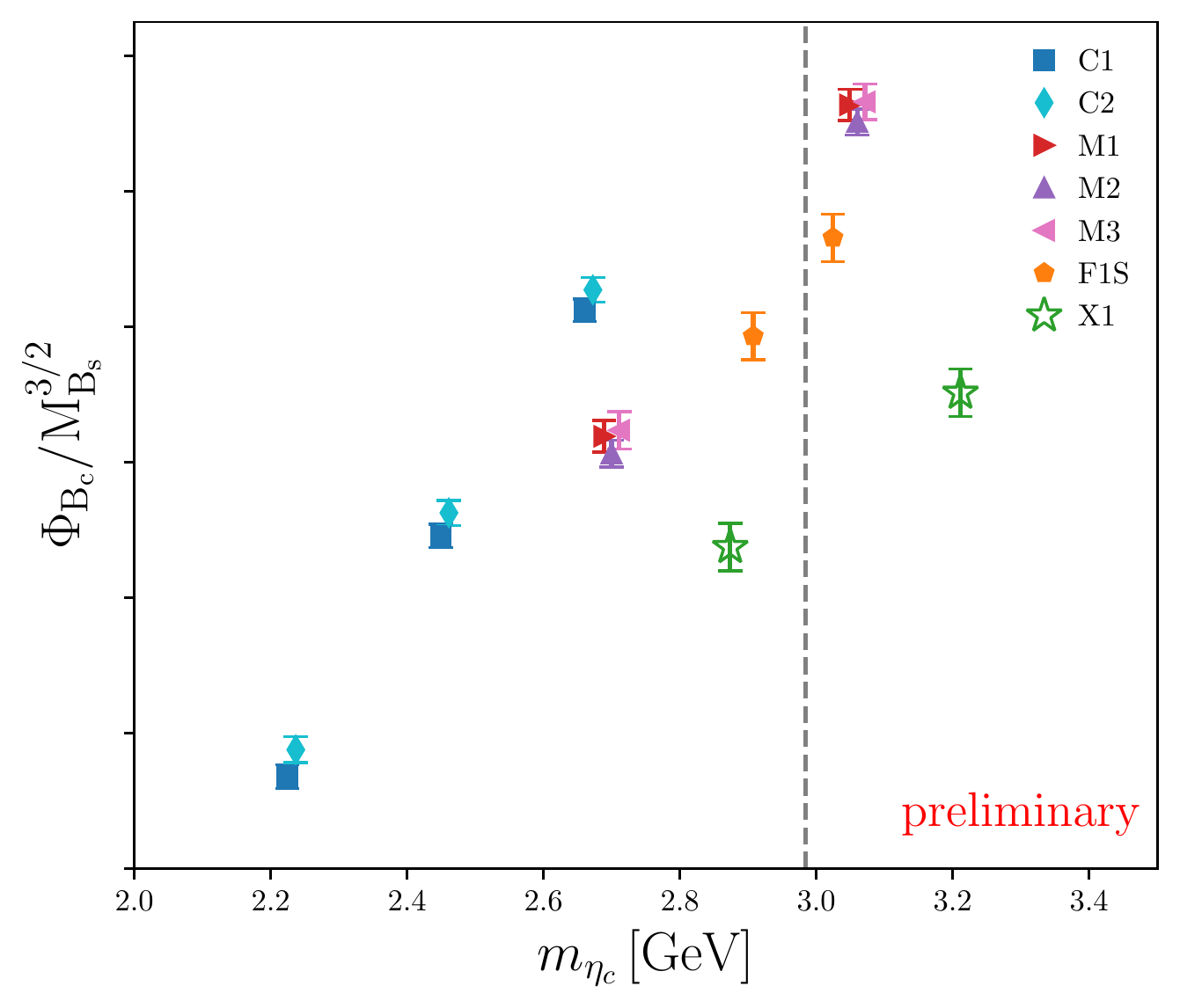}
    \caption{Analysis of the $B_c$ pseudoscalar dimensionless decay amplitudes. {\it Left plot:} determination of the effective mass on F1S for a single charm mass. {\it Right plot:} collection of decay amplitudes on each ensemble for multiple charm quark masses, represented through the $\eta_c(1S)$ mass (excluding disconnected contributions). The open-symbol data are only representative while the X1 RHQ tuning is still in process.}
    \label{fig:BcPhi}
\end{figure}

In analysing the $B_c$ system, one finds a difference from $B$ mesons with light/strange valence quarks in that the heavier $c$ valence quark leads to higher statistical precision in the correlators, given the same number of measurements. 
However, we are instead more sensitive to systematic effects, in particular charm discretisation errors.
In the left plot of Figure~\ref{fig:BcPhi}, one can see that the statistical precision is a factor $\sim2$ higher than previously.

On each ensemble, we extract the decay amplitude at multiple charm quark masses and then on C we extrapolate to the physical charm mass, while on M, F, and X we bracket the physical value and interpolate. 
We define the physical charm point through the $\eta_c(1S)$ meson mass (neglecting disconnected diagrams), matching to the measured value $M_{\eta_c}=2983.9(4)\,$MeV \cite{ParticleDataGroup:2022ynf}.
Until the RHQ tuning is complete and we have finalised data on X1, we do not settle on a fit Ansatz for the combined continuum and physical charm extrapolation. 
In the right plot of Figure~\ref{fig:BcPhi}, we present the pseudoscalar decay amplitudes of the $B_c$ meson on each ensemble.
The different slopes one can see between ensembles indicate cut-off effects which need to be controlled in the final extrapolation. 
We use open symbols for the X1 data to indicate that these are not finalised until the RHQ tuning is finished, and therefore their vertical positioning will change in the final analysis. 
However the slope between the two data points indicates this ensemble will indeed be important in extrapolating to physical values. 

\section{Summary}
$B$ meson decay constants are important quantities for understanding the hadronic physics in weak decays of the SM. 
Here we have presented ongoing work to extract decay constants for both pseudoscalar and vector $B$ mesons with light, strange, and charm valence quarks using the effective RHQ action to simulate physical $b$ quarks.
Analysis of pseudoscalar $f_{B_q}$ for $q=l,s$ including isospin-breaking effects in the light quark is ongoing.
Systematic effects for the $B_c$ meson motivated the inclusion of the X1 ensemble where the nonperturbative tuning of the RHQ action parameters is in progress. 
In addition, we are working on extracting $B^*$ vector meson states and determining their decay constants. 
Once we have collected all measurements, we will obtain renormalised continuum results for all decay constants, as well as interesting ratios such as the SU(3)-breaking estimate $f_{B_s}/f_{B^{0,+}}$ and the vector-to-pseudoscalar ratio $f_{B_{l,s,c}^*}/f_{B_{l,s,c}}$ where an important question remains as to which side of unity this quantity should lie.

\acknowledgments 
\noindent
We thank our RBC/UKQCD colleagues for helpful discussions and suggestions. 
Measurements performed on the X1 ensemble make use of \texttt{Grid}~\cite{GRID,Grid16} and \texttt{Hadrons}~\cite{Hadrons22}.
Computations used resources provided by the USQCD Collaboration, funded by the Office of Science of the US~Department of Energy and by the \href{http://www.archer.ac.uk}{ARCHER} UK National Supercomputing Service, as well as computers at Columbia University, Brookhaven National Laboratory, and the OMNI cluster at the University of Siegen. 
We used gauge field configurations generated on the DiRAC Blue Gene~Q system at the University of Edinburgh, part of the DiRAC Facility, funded by BIS National E-infrastructure grant ST/K000411/1 and STFC grants ST/H008845/1, ST/K005804/1 and ST/K005790/1.  
M.B. is supported by Deutsche Forschungsgemeinschaft (DFG, German Research Foundation) through grant 396021762 - TRR 257 “Particle Physics Phenomenology after the Higgs Discovery”.

\bibliographystyle{JHEP-notitle}
\bibliography{B_meson}

\end{document}